\documentstyle[aps,prl,epsfig]{revtex}

\twocolumn
\newcommand{\Hm}{H_{\rm melt}^c}
\newcommand{\Hc}{H^c}
\newcommand{\Ha}{{H^{ab}}}
\begin{document}
\twocolumn[\hsize\textwidth\columnwidth\hsize\csname@twocolumnfalse\endcsname
\draft
\title{Step-wise Behavior of Vortex-Lattice Melting Transition in Tilted Magnetic Fields
in Single Crystals
Bi$_2$Sr$_2$CaCu$_2$O$_{8+\delta}$}

\author{J. Mirkovi\'c$^{a,b}$, S.E. Savel'ev$^{a,}$\cite{byline1},
E. Sugahara$^a$ and K. Kadowaki$^a$}
\address{$^{a}$Institute of Materials Science, The University of Tsukuba, 1-1-1 Tennodai, Tsukuba 305-8573, Japan, and \\ CREST, Japan Science and Technology Corporation (JST), Japan}

\address{$^b$ Faculty of Sciences, University of Montenegro, PO Box 211, 81000 Podgorica, Montenegro, Yugoslavia }
\maketitle
\begin{abstract}
The vortex lattice melting transition in single crystals $\rm
Bi_2Sr_2CaCu_2O_{8+\delta}$ was studied by the in-plane
resistivity measurements in magnetic fields tilted away from the
$c$-axis to the $ab$-plane.  In order to avoid the surface barrier
effect which hinders the melting transition in the conventional
transport measurements, we used the Corbino geometry of electric
 contacts. For the first time, the complete $\Hc-\Ha$ phase diagram
of the melting-transition in $\rm Bi_2Sr_2CaCu_2O_{8+\delta}$ is
obtained.  The $c$-axis melting field component $\Hm$ exhibits the novel,
step-wise dependence on the in-plane magnetic fields $\Ha$ which is
discussed on the base of the crossing vortex lattice structure.  The sharp
change of resistance behavior observed near the $ab$-plane suggests
transformation from first-order to second-order phase transition.
\end{abstract}
\tighten

\vskip.2pc]

\narrowtext

After discovery of high temperature superconductors, the phase
diagram of the vortex state has been reconsidered because the
novel vortex phases and phase transitions have been recognized
\cite{blatter}. Among them, the most intriguing phenomena is the
first-order vortex-lattice melting transition (hereafter VLMT)
\cite{nelson}, \cite{VLMT}.  However, despite a lot of
experimental and theoretical efforts, the nature of this
phenomenon is not completely understood yet. In particular, it is
not clear what happens with melting transition in highly
anisotropic systems such as $\rm Bi_2Sr_2CaCu_2O_{8+\delta}$ if
magnetic field is tilted away from the $c$-axis, especially close
to the $ab$-plane.

 In contrast to melting transition in
YBa$_2$Cu$_3$O$_{7-\delta}$ \cite{schiling} described well by the 
anisotropic 3D Ginzburg-Landau (GL) theory \cite{blater3d}, 
it was noticed \cite{konc} that VLMT in 
$\rm Bi_2Sr_2CaCu_2O_{8+\delta}$ 
obeys neither 3DGL nor 2D \cite{kes} scaling. 
Later, Oii {\it et al.} \cite{oii} found that
the $c$-axis melting field component $\Hm$ depends linearly on the 
in-plane magnetic
field $\Ha$ in Bi$_2$Sr$_2$CaCu$_2$O$_{8+\delta}$. The authors
explained the observed behavior through suppression of the
Josephson coupling by the in-plane magnetic fields. They concluded
that the anisotropy increases with $\Ha$ and should become
infinite in a certain magnetic field at which the decoupling of
the vortex lattice occurs. On the other hand, Koshelev \cite{koshelev}
interpreted the linear dependence of $\Hm$ on $\Ha$ as an
indication of the crossing lattice of Josephson vortices (JVs) and
pancake vortex stacks (PVSs). According to this model, the linear
dependence  $\Hm(\Ha)$ breaks down and transforms into plateau as
soon as the JV cores overlap.  Such plateau-like behavior
was observed recently \cite{lt22,koncconmat},  but there is still
a question about the underlying mechanisms governing the melting
transition in the tilted magnetic fields.  In this regard, it is
challenging to investigate what happens to the melting transition in
much higher in-plane magnetic fields.

Here, we have investigated the melting transition using the in-plane
resistivity measurements in the oblique magnetic fields in the
whole angular range and found the dramatic new features,  beyond
pre-existing experimental data and theoretical models. The
 $\Hc-\Ha$ phase diagram of melting transition exhibits the peculiar step-wise behavior.
 Besides, the transformation from the
first-order to the second-order phase transition near $ab$-plane 
is indicated.

It is worth to note that the novel experimental findings were
obtained essentially by using the unconventional Corbino
geometry of electric contacts in the resistivity measurements.
As it is  known well, the transport measurements in the conventional
four probe strip geometry  in $\rm Bi_2Sr_2CaCu_2O_{8+\delta}$
\cite{fuchs,mir} do  not probe the true bulk properties
of superconductor because  surface barriers short-cut the current
path and, as a consequence, the currents flow only near the sample edges.
Therefore, the features of VLMT are also strongly hindered in
resistivity measurements in the platelet samples.  In order to
avoid the smearing effect of VLMT and the nonlinear behavior in the vortex 
liquid
phase due to surface barriers \cite{lt22_2} we have used the Corbino electric
contact geometry at which currents flow radially, far from the
edges of sample (see inset of Fig. 1).  Such experimental
set-up considerably improved quality of resistivity data compared
with conventional technique.

The in-plane resistivity measurements were performed for two
as-grown single crystals \cite{kadgrowth} $\rm
Bi_2Sr_2CaCu_2O_{8+\delta}$ with transition temperature $T_c=
90.3$ K and $T_c= 90.0$ K for samples \#S1 and \#S2, respectively.
The diameters of the Corbino discs were $D=1.9$ mm  for \#S1 and 2.7 mm
 for \#S2, while the thickness was $t\approx 20\ \mu$m for both
samples. The resistance was measured by two pairs of electric
contacts  using the ac lock-in method at a low frequency of 37 Hz.
They agree with each other in a geometrical configuration error of
about 5\%. The measurements of resistance were carried out as a
function of the magnetic field $H$ at its different orientation
$\theta$ with respect to the $c$-axis at various temperatures. The
magnetic field was rotated by using 70 kOe split coil with fine
goniometer with angular resolution of $0.001^\circ$.
\begin{figure}[btp]
\begin{center}\leavevmode
\includegraphics[width=1.15\linewidth]{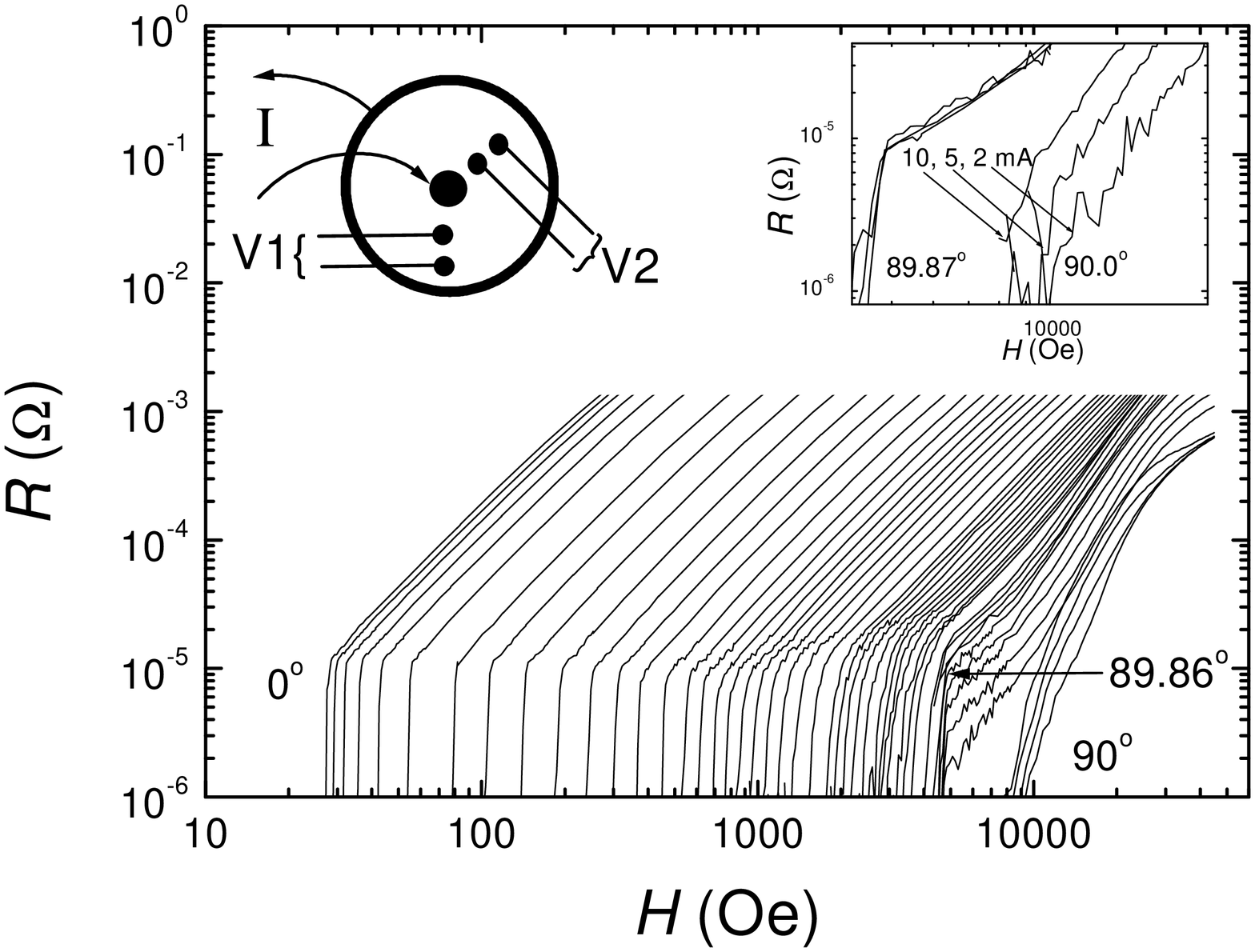}
\caption{The magnetic field dependence of the resistance measured
at $T=85.2$ K at various field orientations with respect to the
$c$-axis in sample \#S1 with the current of 5 mA. Left inset: a
sketch of the sample with electric contacts in the Corbino
geometry. Right inset: the magnetic field dependence of the
resistance measured for two field orientations by three current levels.}
\label{f1}
\end{center}
\end{figure}

The magnetic field dependence of the resistance at different
orientations, which was measured at a particular temperature of 85.2 K, is
shown in Fig. 1.  The steep resistance drop attributed to
VLMT \cite{watauchi,prljm} is clearly detected even in oblique magnetic
fields and, surprisingly, changes only weakly up to $\theta=89.86^\circ$.
We should emphasize that the step height of the resistance anomaly 
practically remains the same  and sharply separates the vortex solid 
 and vortex liquid phases. As far as we know, 
 the VLMT has never been found so close to the $ab$-plane. However, as the
angle $\theta$ exceeds the critical value, only 0.14$^\circ$
away from the $ab$-plane, the resistance level of the  kink
suddenly begins to decrease, while the sharpness of the anomaly  gets even stronger.
 Then, the kink feature finally vanishes in the instrumental noise, 
 and qualitatively new, smooth dependence $R(H)$ sets in. 
 Instead of the quadratic field dependence of the
resistance, found  in the  vortex liquid phase, $R(H)$ follows  
essentially the non-power law in the
narrow angle region near the $ab$-plane. 
In addition, it is remarkable that the {\it Ohmic}
response  of the resistance observed in the vortex liquid state at
angles $\theta<89.96^\circ$ is replaced by the {\it non-Ohmic} behavior 
(see inset of Fig. 1). 
 This peculiar resistance behavior
may indicate the novel vortex phase in magnetic fields parallel to
the $ab$-plane and the change of character of phase transition
from first-order to second-order. Such experimental finding 
could be related
to the scenario proposed earlier \cite{nelson1}, which suggests
transformation from the first-order vortex lattice melting
transition into the second-order vortex-lattice ---vortex-smectic
phase transition in magnetic fields near the $ab$-plane.
\begin{figure}[btp]
\begin{center}\leavevmode
\includegraphics[width=1.15\linewidth]{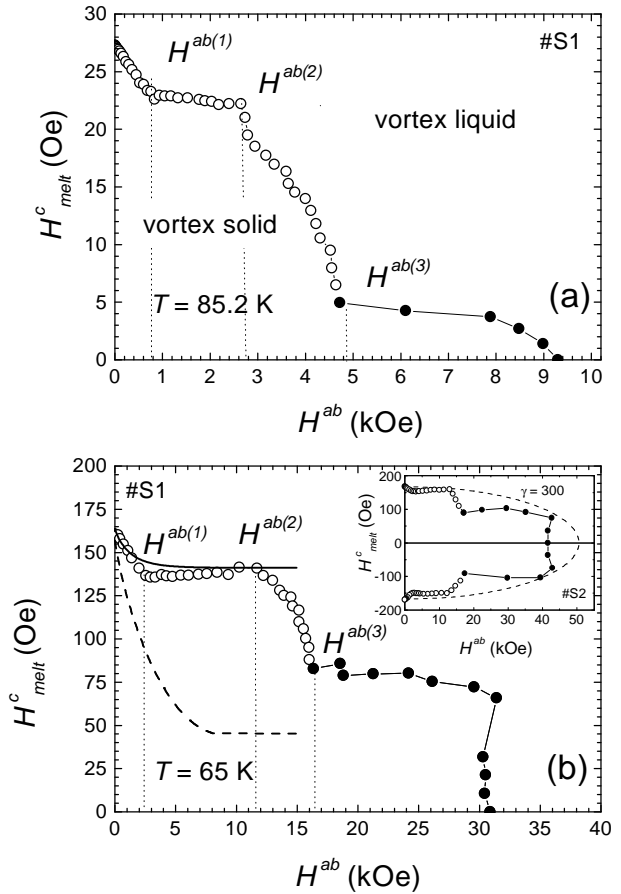}
\caption{Vortex lattice melting transition in the $H^c-\Ha$ plane
for the sample \#S1 at two different temperatures of 85.2 K (a)
and 65 K (b). Dashed line in (b) is the fitting curve obtained from
eq. (8) in [9]. The solid line is the fitting curve
calculated after taking into account the interaction between PVs
and currents of JVs. Inset in (b): $H^c-\Ha$ phase diagram for
sample \#S2 at $T=62$ K. The solid line corresponds
to the anisotropic 3D GL model with the anisotropy parameter of
$\gamma=300$.}
\label{f2}
\end{center}
\end{figure}

Using the resistance data obtained at two temperatures of 85.2 K
and 65 K for the sample \#S1,  the VLMT phase diagram is constructed in the
$H^c$--$\Ha$  plane (Fig. 2).  The melting transition was defined
by the resistance criterion $R=10^{-6}\ \Omega$ in the whole
angular range. The in-plane field dependence of the $c$-axis
melting field component $\Hm$ exhibits a very intriguing step-wise
behavior at both
 temperatures. At first stage, $\Hm$ decreases linearly with
 the in-plane magnetic field $\Ha$.
Then, the linear dependence is abruptly terminated at field
$\Ha^{(1)}$ and transforms into plateau which continues to the
field $\Ha^{(2)}$ where, surprisingly, the $c$-axis component of
the melting field $\Hm$ starts to decrease again. This finding is
beyond the pre-existing experimental data and theoretical models.
Finally, in the in-plane magnetic fields above $\Ha^{(3)}$, the
transition field $\Hm$ exhibits the second weak in-plane field
dependence without the resistance kink anomaly (this region is
indicated by filled symbols in Fig. 2). The almost identical
 step-wise behavior of the melting transition is found also in the
sample \#S2, as shown in the inset of Fig. 2b at somewhat lower
 temperature of 62 K \cite{magnetization}.

The observed in-plane field dependence of the melting transition
is in the strong contrast with the 2D scaling \cite{kes} ruled
only by the out-of-plane magnetic field component.  The inset in
Fig. 2b demonstrates also the essential discrepancy between our
data and the fitting curve based on the 3D scaling
\cite{blater3d} $H_{\rm melt}(\theta)=H_{\rm
melt}(0)/(\cos^2\theta+ \gamma^{-2}\sin^2\theta)^{1/2}$, where
$H_{\rm melt}(0)$ is the melting field for $\theta=0^\circ$ and
$\gamma=300$ is chosen as the anisotropy parameter. 
 Furthermore,
the second decrease of the field dependence $\Hm(\Ha)$ after
plateau can not be explained in the frame of the phenomenological
concept of ''decoupled'' vortex state proposed by Ooi {\it et
al.}\cite{oii}, since there would not be any in-plane field
dependence of the melting transition above the decoupling field.

The first theoretical explanation  of the unusual linear
dependence  $\Hm(\Ha)$ was given by Koshelev  \cite{koshelev} who
analyzed the crossing vortex lattice with interaction between 
pancake vortices (PVs)
and JVs. Using the general thermodynamical equality of the free
energies of vortex solid and vortex liquid phases at the melting
transition, Koshelev obtained the linear dependence of the
$c$-axis melting field component:  $\Hm(\Ha)=\Hm(0) -4\pi
\epsilon_J\Ha/(\Delta B\Phi_0)$, where $\epsilon_J$ is the energy
of a JV in the presence of PV lattice, $\Delta B$ is the jump of the
magnetic induction at the melting point, and $\Phi_0$ is the
quantum of the magnetic flux \cite{termody}. In general, the
energy of JV contains both, the self energy of JV,
$\epsilon_J^{self}$  (eq. (3) in \cite{koshelev}), and the
interaction energy $E_{PJ}$ of PVS with currents generated by JV,
{\it i.e.} $\epsilon_J=\epsilon_J^{self}-E_{PJ}$. The self energy
contribution elevates the free energy of vortex solid phase and
reduces the stability of the vortex crystal, while
 the second term acts in the opposite way, through decreasing
 of the free energy.  The latter, interaction term, $E_{PJ}$ was omitted
 in the analysis \cite{koshelev} of the melting transition for the case
 of a dense PV lattice. Figure 2b
shows the fitting curve (dashed line) obtained from the equation
(8) of Reference \cite{koshelev} with rather high value of
anisotropy parameter $\gamma=1500$ (the lower values of $\gamma$
give even stronger discrepancy with experimental data), the in-plane penetration 
depth $\lambda=2000/\sqrt{1-T^2/T_c^2}\ {\rm \AA}$, the interlayer 
distance $s=15\ {\rm \AA}$, and $\Delta B=0.35$ Oe. It is easy
to see that the
 above mentioned model qualitatively describes our experimental data but
the calculated average slope ($d\Hm/d\Ha\approx 0.025$) as well as
the expected breaking field ($\Ha^{(1)}\approx 7$ kOe) appeared to
be several times higher than our experimental values (
$d\Hm/d\Ha\approx 0.01$, $\Ha^{(1)}\approx 2.5$ kOe).
\begin{figure}[btp]
\begin{center}\leavevmode
\centerline{\includegraphics[width=1.15\linewidth]{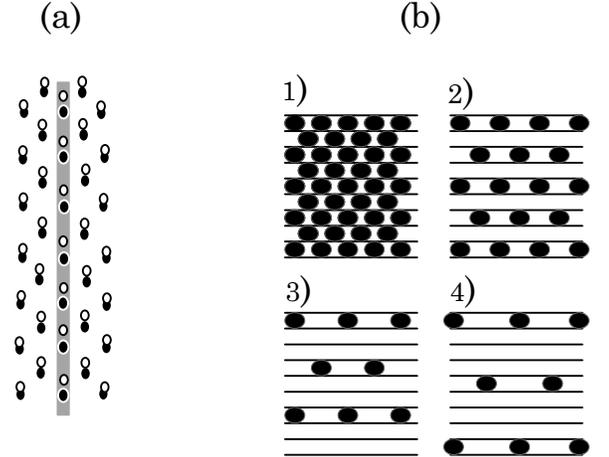}}
\caption{a) The sketch of the PV lattice deformation 
 due to the
interaction between PVs and currents of JVs: filled circles correspond to the
positions of PVs in an ideal lattice, open circles mark
  PVs in the presence of JV. b) Schematic pictures of
the JV lattice with different periods along the $c$-axis.}
\label{f3}
\end{center}
\end{figure}

On the other hand, in the experimentally studied magnetic field
range and for the more realistic value of the anisotropic parameter 
$\gamma\approx 300$, the
size of the nonlinear JV core $\lambda_J=\gamma s$ 
is about the same distance as the distance between
pancakes vortices, $a_p$. In such a  case, according to Koshelev
\cite{koshelev}, the interaction between PVSs and currents
generated by JVs becomes important, and it is necessary to
consider also the energy $E_{PJ}$.  Due to Lorentz force,  PVs shift from
their equilibrium positions (Fig.~3 a) reducing the free energy of
the solid phase.  Thus, the vortex crystal is stabilized and the
$c$-axis melting field component decreases more slowly with the in-plane
magnetic field. Therefore, the transformation of the linear dependence
$\Hm(\Ha)$ into plateau could be related not to the overlapping of
the JV cores, but to the compensation of the self energy of JVs by
their interaction energy with PVSs. The rough calculations \cite{savelev},
 taking  into account the pinning of PVSs born by the 
 interaction with currents of JVs, improve the correlation between model
 \cite{koshelev}
 and experimental data (Fig. 2b (solid line)).
 The fitting curve was obtained under  assumption that the
pinning term is suppressed logarithmically by the in-plane
field similar to the self-energy of Josephson vortex
\cite{koshelev}. Nevertheless, this term decreases faster in
higher in-plane fields due to diminishing of currents generated by
JVs, when the
distance $b_J$ between JVs along the $c$-axis approaches $2s$.
This means that the mentioned compensation can be destroyed in the
field close to $\Phi_0/b_Ja_J=\Phi_0/2\sqrt{3}\gamma s^2$ 
($a_J$ is the distance between JVs in a layer)
and the new decay of
the $c$-axis melting field component could follow again; for
$\gamma=300$, $\Phi_0/2\sqrt{3}\gamma s^2\approx 9$ kOe, which is 
close to the observed value $\Ha^{(2)}$ at $T=65$ K.  
Next, we point out that the disappearance of
the kink anomaly at the field $\Ha^{(3)}$ could indicate the
so-called ''lock-in'' transition \cite{fein} above which the
observed behavior could be related to the melting of JV lattice
via the second-order phase transition \cite{nelson1,hu}.

Finally, we  note that the unusual step-wise behavior of the
melting transition seems to be induced by the layer structure of
Bi$_2$Sr$_2$CaCu$_2$O$_{8+\delta}$ which governs, in particular,
the behavior of the JV lattice. According to Bulaevskii and Clem
\cite{bul}, the JV lattice undergoes the sequential structural
first-order phase transitions between lattices with different
periods.   These transitions are based on the fact that the JVs
are able to occupy only the space in-between the CuO$_2$ planes,
{\it i.e.} the distance $b_J$ between JVs along the
$c$-axis should have discrete values $b_J=ks$, where $k$ is an
integer. The structural transitions  could explain the sharpness
of dependence $\Hm(\Ha)$ at points $\Ha^{(1)}$ and $\Ha^{(2)}$
since  the interaction between PVS and JVs is sensitive to the
change of JV lattice parameters $b_J$ and $a_J$.  
Interestingly, the ratio of the magnetic
fields where last two structural transitions happen
 (from the lattice 1 to 2 and from 2 to 3 in Fig. 3b) is 25/9,
 being close to the experimentally observed value
 $\Ha^{(2)}/\Ha^{(3)}\approx 3-4$ at all temperatures.

In summary, we present for the first time the complete $\Hm-\Ha$
phase diagram of VLMT in the single crystals
Bi$_2$Sr$_2$CaCu$_2$O$_{8+\delta}$. In strong contrast to the
conventional superconductors and YBa$_2$Cu$_3$O$_{7-\delta}$, 
we observe the step-wise behavior of
the melting transition, which reflects the layer structure of the
system and could be related to interaction between pancake
vortices and Josephson vortex lattice. The resistivity behavior
indicates that melting transition changes its character from
first-order to second-order phase transition in magnetic fields
applied very close to the $ab$-plane.

We appreciate the enlightening discussions with A.E.
Koshelev, J.R. Clem, M. Tachiki, G. Crabtree and  A.I. Buzdin.


\begin{references}
\bibitem{magnetization} We emphasize that our
experimental results agree well to ac
magnetization measurements over the common measured
interval (for $\Ha<\Ha^{(2)}$ compare \cite{lt22} and
\cite{koncconmat}). However, in contrast to ac and dc
magnetization technique \cite{konc,oii,koncconmat,kkkk}, the
resistance measurements with Corbino geometry do not lose
sensitivity for fields aligned near the $ab$-plane, and thus, make
it possible to get the complete $\Hm-\Ha$ phase diagram in the
whole angular range.
\end{references}
\end{document}